\documentclass[10pt,a4paper,twocolumn]{article}
\pdfoutput=1
\usepackage{amsmath,amssymb,graphicx,bm}
\usepackage[top=1in, bottom=1.0in, left=0.5in, right=0.5in]{geometry}

\usepackage{times}
\usepackage{lettrine}
\usepackage[usenames,dvipsnames]{color}

\usepackage{caption}
\DeclareCaptionLabelSeparator{bar}{\ \textbar\ }
\usepackage[super,comma,sort&compress]{natbib}
\usepackage{doi}
\usepackage{hyperref}
\hypersetup{%
pdftitle={High-temperature superfluidity with indirect excitons in van der Waals heterostructures},%
pdfauthor={M. M. Fogler et al.},%
pdfpagemode={UseNone},%
pdfstartview={FitH},%
breaklinks=true,%
citecolor=blue,%
colorlinks=true,%
linkcolor=blue,%
urlcolor=blue}

\begin{document}
\title{High-temperature superfluidity with indirect excitons in van der Waals heterostructures}

\author{M. M. Fogler,$^1$ L. V. Butov$^1$ \& K. S. Novoselov$^2$}

\twocolumn[
\maketitle

\parbox{0.75\textwidth}{
All known superfluid and superconducting states of condensed matter are enabled by composite bosons (atoms, molecules, Cooper pairs) made of an even number of fermions. Temperatures where such macroscopic quantum phenomena occur are limited by the lesser of the binding energy and the degeneracy temperature of the bosons. High critical temperature cuprate superconductors set the present record of $\sim100\,\mathrm{K}$. Here we propose a design for artificially structured materials to rival this record. The main elements of the structure are two monolayers of a transition metal dichalcogenide
separated by an atomically thin spacer.
Electrons and holes generated in the system would accumulate in the opposite monolayers and form bosonic bound states --- the indirect excitons. The resultant degenerate Bose gas of indirect excitons would
exhibit macroscopic occupation of a quantum state and vanishing viscosity at high temperatures.
}

\vspace{5.00in}

{\small
\textcolor{Orange}{\hrulefill}

$^1$Department of Physics,
University of California San Diego, La Jolla, 9500 Gilman Drive,
California 92093, USA.
$^2$School of Physics and Astronomy,
University of Manchester, Manchester M13 9PL, UK.
}
]

\twocolumn[]

\lettrine[lines=3, lraise=0.1, nindent=0.0em, findent=0.2em]{\textcolor{Orange}{C}}{}
oherent states of excitons have been a subject of intense theoretical studies.~\cite{Keldysh1968cpe, Lozovik1976nms, Fukuzawa1990pcl}
A general framework for creation and manipulation of degenerate gases of indirect excitons has been established in prior experimental studies of GaAs-based coupled quantum wells (CQW) where electrons and holes are confined in GaAs quantum wells separated by a thin AlGaAs barrier.~\cite{High2008cef, High2012sci} Here we apply similar principles to the design of a CQW from atomically thin materials stacked on top of each other. Research on such van der Waals heterostructures is gaining momentum in the last few years,~\cite{Geim2013vdw} and their quality and availability is steadily improving.

\begin{figure}[h]
\begin{center}
\includegraphics[width=3.0in]{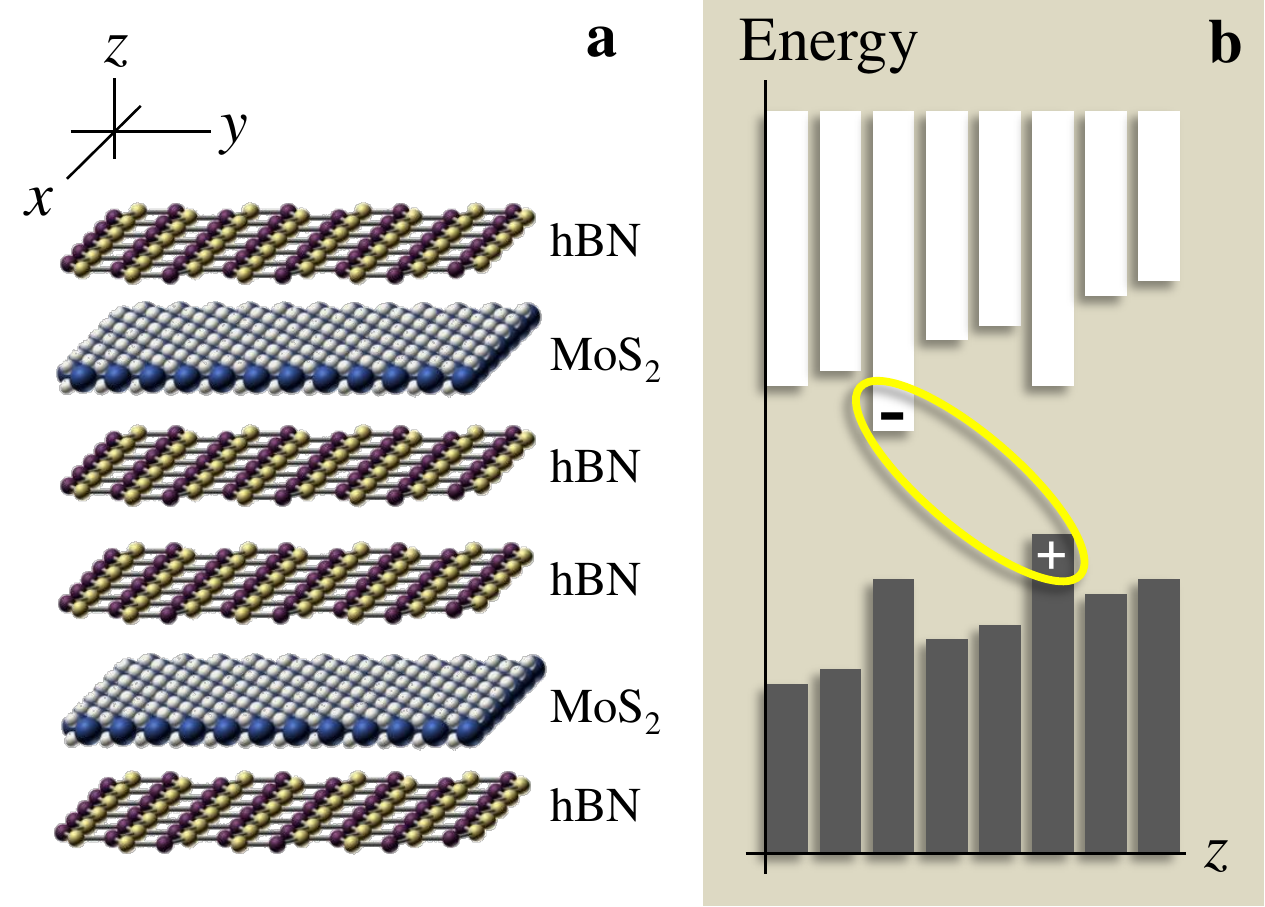}
\end{center}
\caption{\textbf{Schematics of the proposed device.} \textbf{(a)} geometry \textbf{(b)} band structure. The ellipse indicates an indirect exciton composed of an electron ($-$) and a hole ($+$).
}
\label{fig:device}
\end{figure}

In the proposed device (Fig.~\ref{fig:device}a), an indirect exciton is composed from an electron and a hole located in two different MoS$_2$ layers separated by an hexagonal boron nitride (hBN) insulating barrier and surrounded by hBN cladding layers. The $z$-direction electric field is controlled by voltage applied to external electrodes. The applied field modifies the band structure in a way that it becomes advantageous for optically excited electrons and holes to reside in the opposite MoS$_2$ monolayers and form indirect excitons (Fig.~\ref{fig:device}b).

The design of the MoS$_2$/hBN structure is similar to that of GaAs/AlGaAs CQW we studied previously,~\cite{High2008cef, High2012sci} except GaAs is replaced by MoS$_2$ and AlGaAs by hBN.
Here we predict the phase diagram of such a device
using the results of numerical calculations and scaling arguments.
Our most intriguing finding is that in the proposed MoS$_2$/hBN structures degenerate Bose gas of indirect excitons can be realized at record-high temperatures.

\subsection*{Results}

\paragraph{Degeneracy temperature.}
The characteristic temperature $T_d$ at which excitons become degenerate is determined by their density $n_x$ per flavor (spin and valley), and effective mass $m_x = m_e + m_h$:
\begin{equation}
k T_d = \frac{2\pi\hbar^2}{m_x}\, n_x\,.
\label{eqn:T_d}
\end{equation}
Here $m_e$ and $m_h$ are the electron and hole effective masses. For $n_x \sim 10^{10}\,\mathrm{cm}^{-2}$, and $m_e = 0.07$, $m_h = 0.15$, $m_x = 0.22$
representative of GaAs CQW, we find $T_d \sim 3\,\mathrm{K}$.
(Here and below all the masses are in units of the bare electron mass.)
At such temperatures long-range spontaneous coherence of indirect excitons is observed.~\cite{High2012sci}

%\noindent{}
%%%%% FIG %%%%%
%%%%%%%%%%%%%%%

To explain why MoS$_2$-based CQW would possess much higher $T_d$ than GaAs-based ones we consider characteristic length and energy scales in the problem, which are the exciton Bohr radius $a_x = {\hbar^2 \epsilon_1}\,/\,{\mu e^2}$ and the Rydberg energy $\mathrm{Ry}_x = {\hbar^2}\,/\, {2\mu a_x^2}$ defined in terms of reduced mass $\mu = m_e m_h / m_x$.
For estimates we use the effective static dielectric constant $\epsilon_1 = 4.9$ of hBN.
We also use the calculated value $\mu = 0.25$~\cite{Xiao2011csv, Berkelbach2013tnc} for MoS$_2$.
Based on measurements~\cite{Zhang2013dot} done on a related compound MoSe$_2$, this value should be accurate to within 20\%.
We find $a_x \approx 1.0\,\mathrm{nm}$, much shorter than $15\,\mathrm{nm}$ in GaAs CQW.
In turn, $\mathrm{Ry}_x \approx 140\,\mathrm{meV}$ in the proposed device, about forty times larger than in GaAs.

\begin{figure}[h]
\begin{center}
\includegraphics[width=2.8in]{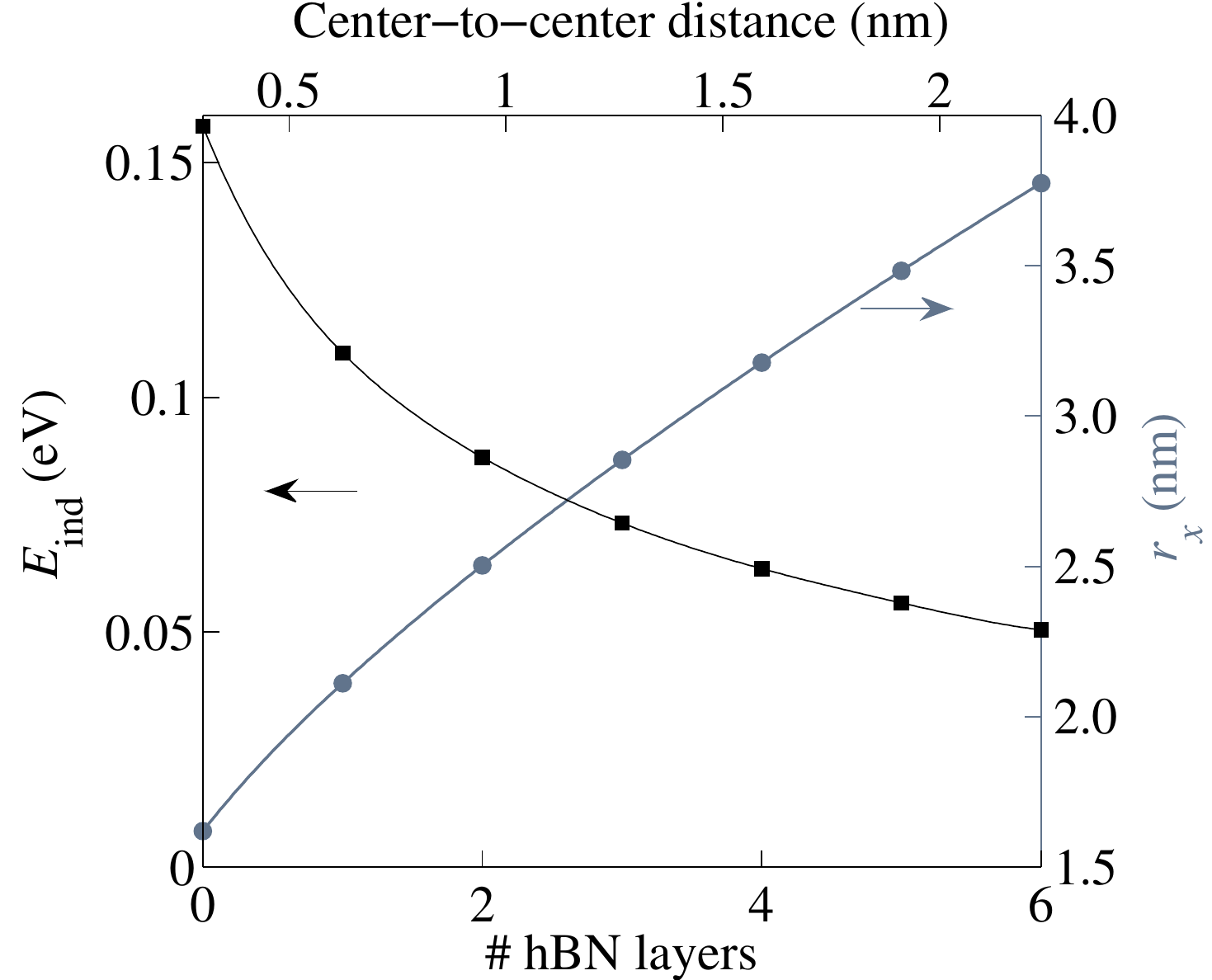}
\end{center}
\caption{\textbf{Estimated exciton parameters.} $E_{\mathrm{ind}}$ (left axis) is the binding energy and $r_x$ [right axis, Eq.~\eqref{eqn:r_x}] is the in-plane gyration radius of indirect excitons.
The bottom axis is the number of the hBN spacer layers and the top axis is the corresponding center-to-center distance between the MoS$_2$ layers in the heterostructure.
}
\label{fig:E_ind}
\end{figure}

Since it enters the denominator of Eq.~\eqref{eqn:T_d},
one may think that having larger $m_x$ is unfavorable for attaining higher $T_d$. In fact, when $n_x$ can be controlled, the opposite is true.
Indeed, Eq.~\eqref{eqn:T_d} can be rewritten as
\begin{equation}
{k T_d}\, = {4\pi} ({m_e m_h} / {m_x^2}) (n_x a_x^2) \, {\mathrm{Ry}_x}.
\label{eqn:T_d_atomic}
\end{equation}
The upper limit on $n_x a_x^2$ is imposed by quantum dissociation of excitons that occurs when the ratio of the exciton size (the in-plane gyration radius) $r_x$ and the mean 
inter-exciton distance $1 / \sqrt{n_x}$ reaches the critical value~\cite{DePalo2002eci, Schleede2012pdb, Maezono2013eab} of about $0.3$.
In the case of our primary interest where interlayer center-to-center distance $c \approx a_x$ and $m_e = m_h = 2 \mu$,
we estimate $r_x = 2.4 a_x$ (Fig.~\ref{fig:E_ind}),
so that the corresponding Mott critical density $n_{\mathrm{M}}$ is set by the condition
\begin{equation}
n_{\mathrm{M}} a_x^2 \sim 0.02\,.
\label{eqn:n_M}
\end{equation}
Substituting this into Eq.~\eqref{eqn:T_d_atomic}, we obtain
\begin{equation}
{k T_d}^{\mathrm{max}} \sim 0.06\, {\mathrm{Ry}_x}\,.
\label{eqn:T_d_max}
\end{equation}
Hence, the key to high $T_d$ is the enhanced value of $\mathrm{Ry}_x$. The exciton binding energy $E_\mathrm{ind}\approx 0.6\, \mathrm{Ry}_x$ (Fig.~\ref{fig:E_ind}) does not pose further fundamental limitations.
Actually, the conclusion that the theoretical maximum of $T_d$ is proportional to $\mathrm{Ry}_x$ follows from dimensional analysis. If $c \sim a_x$, $m_e \sim m_h$, and excitons are treated as an equilibrium Bose gas, then $\mathrm{Ry}_x$ is the only relevant energy scale in the problem.

%%%%% FIG %%%%%
%%%%%%%%%%%%%%%

In reality, only quasi-equilibrium state is possible because of exciton recombination, and so limitations posed by the finite exciton lifetime $\tau$ must be discussed.
Radiative recombination of indirect excitons requires interlayer tunneling. The rate of this process can be made exponentially small by adjusting the barrier width. In GaAs CQW $\tau$ can be routinely made in the range of $\mu$s and can be controllably varied over several orders of magnitude. The large $\tau$ enables creation of indirect exciton gases of high density $n_x = P \tau$ with low optical excitation power $P$ that does not cause overheating of either the lattice or the excitons. This is why for the task of achieving cold dense exciton gases, indirect excitons are superior to conventional bulk excitons or two-dimensional (2D) direct excitons (bound states of electrons and holes in the same layer).

In the proposed device, the interlayer tunneling rate would decay exponentially with the number $N$ of layers in the hBN spacer.
For $N = 2$ the tunneling rate is comparable to that in the GaAs CQW~\cite{High2008cef, High2012sci} (see Methods) indicating that long-life indirect excitons can be realized in MoS$_2$/hBN structures. The tunneling-limited lifetime can be enhanced by possible rotational misalignment of the MoS$_2$ layers, which makes the excitons indirect not only in real but also in momentum space, similar to excitons in MoS$_2$ multilayers.~\cite{Mak2010atm, Zhang2013dot} Hence, sufficiently long $\tau$ may perhaps be achieved with a monolayer hBN spacer.
Working with very small $N$, one has to worry about possibility of dielectric breakdown of the hBN spacer. To make the indirect exciton more energetically favorable than the direct one, a voltage equal or larger than $(E_{\mathrm{dir}} - E_{\mathrm{ind}}) / e$ must be applied between the MoS$_2$ layers. Assuming the direct exciton binding energy $E_{\mathrm{dir}}$ of approximately \cite{Berkelbach2013tnc, Chernikov2014nhe} $0.5\,\mathrm{eV}$, for $N = 2$ hBN spacer, the required voltage is about $0.4\,\mathrm{V}$, which is safely below the breakdown limit.~\cite{Britnell2012fet}

\begin{figure}[h]
\begin{center}
\includegraphics[width=2.80in]{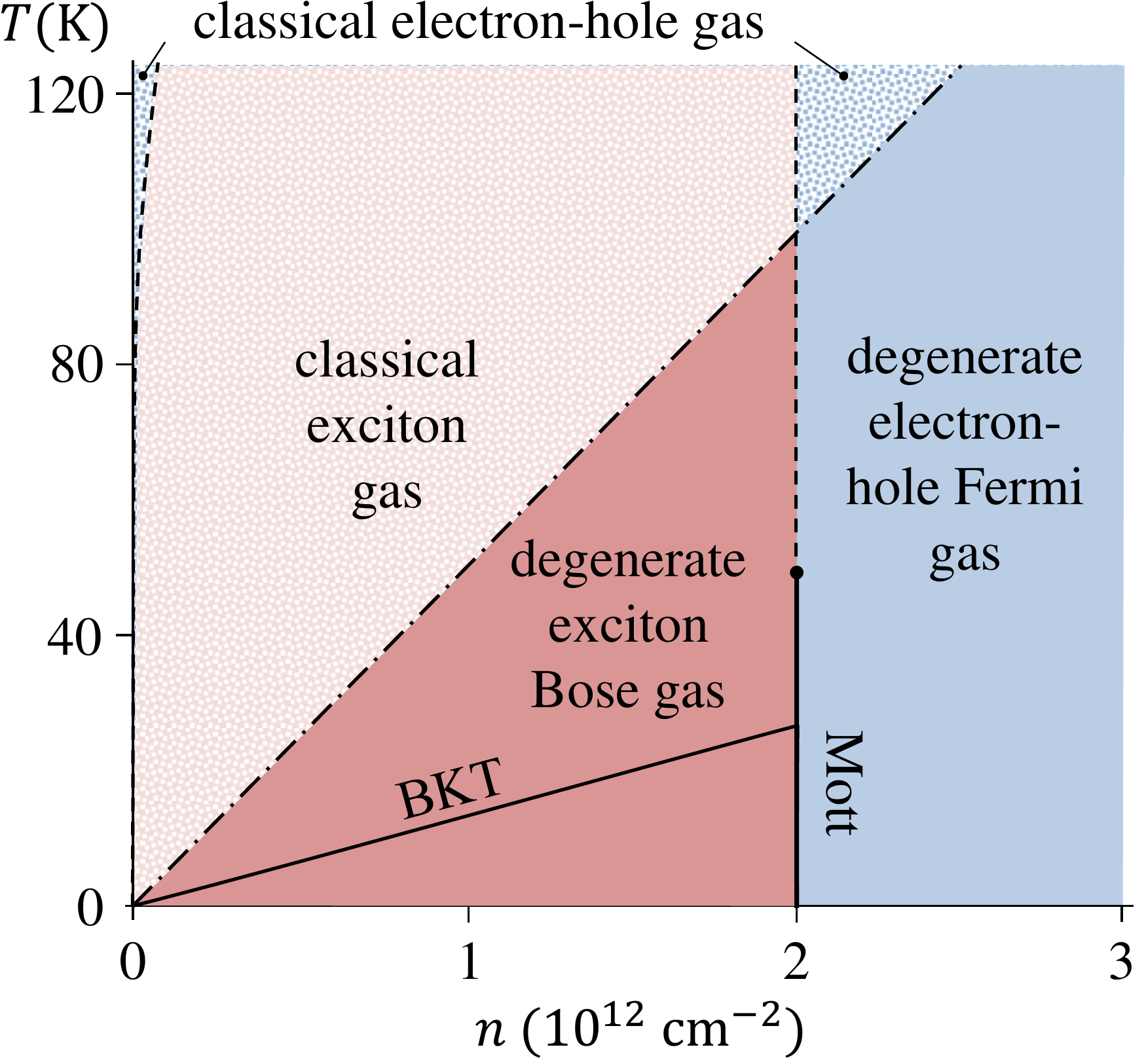}
\end{center}
\caption{\textbf{The phase diagram of the system studied.} $T$ is the temperature and $n$ is the electron density per flavor.
The dashed lines indicate crossovers and the solid lines mark the phase transitions.
The dot is the Mott critical point.
}
\label{fig:T}
\end{figure}

A schematic phase diagram of a neutral electron-hole system in the proposed heterostructure is shown in Fig.~\ref{fig:T}. The solid line represents the Mott transition. It emanates from the $T = 0$ critical point computed according to Eq.~\eqref{eqn:n_M}. At the transition there is a discontinuous jump in the degree of exciton ionization.~\cite{Nikolaev2004mts}
Above the Mott critical temperature $T_{\mathrm{M}}$ the transition changes to a smooth crossover.
Such a crossover has been studied experimentally in
photoexcited single-well GaAs and InGaAs structures.~\cite{Butov1991msm, Kappei2005dom}
We assume that $T_{\mathrm{M}}$ should be lower than $T_d^{\mathrm{max}}$ and use
$k T_{\mathrm{M}} = 0.5 k T_d^{\mathrm{max}} \sim 0.03\,\mathrm{Ry}_x$ in Fig.~\ref{fig:T} for illustration.
The quantum degeneracy line [Eq.~\eqref{eqn:T_d}] shown by the dashed-dotted line in Fig.~\ref{fig:T} demarcates
a crossover from classical to degenerate Bose gas of excitons.
At $n = n_{\mathrm{M}}$ this line extrapolates to the temperature $k T_d^{\mathrm{max}}$ [Eq.~\eqref{eqn:T_d_max}].

Whereas the Mott transition represents quantum dissociation of excitons, thermal dissociation occurs above
the Saha temperature $k T_{\mathrm{S}} \sim ({\pi \hbar^2 n_x}/{m_x}) e^{E_{\mathrm{ind}} / k T_{\mathrm{S}}}$,
which is shown by the dashed line near the $T$-axis in Fig.~\ref{fig:T}.
This line marks a crossover from the exciton phase to a classical plasma.
Since their binding energy is very large,
indirect excitons in MoS$_2$/hBN structure remain stable against the thermal dissociation well above the room temperature in a broad range of electron densities.

Formation of quantum degenerate Bose gas of long-lifetime repulsively interacting indirect excitons leads to local exciton superfluidity below $T_d$.
The superfluidity spreads over a macroscopic area at the Berezinskii-Kosterlitz-Thouless transition temperature \cite{Filinov2010bkt}
\begin{equation}
k T_{\mathrm{BKT}} \approx 1.3\, \frac{\hbar^2 n_x}{m_x}\,.
\label{eqn:T_BKT}
\end{equation}
Superfluid transport of neutral indirect excitons produces dissipationless charge currents in the opposite directions in the two layers.~\cite{Lozovik1976nms}
To observe and utilize this effect, one can, e.g., make separate contacts to each layer and form a closed circuit for the hole layer.
The electric current in the electron layer will then be dissipationless.
This is referred to as the counterflow superconductivity.~\cite{Su2008htm}

The most intriguing conclusion we draw from Fig.~\ref{fig:T} is that in the proposed MoS$_2$/hBN structures degenerate Bose gas of indirect excitons can be realized at record-high temperatures.

%%%%% FIG %%%%%

\subsection*{Discussion}

In the remainder of this paper, we overview phenomena analogous to a coherent state of indirect excitons in other 2D systems.
We choose not to survey systems where electrons and holes reside in the same layer,
such as quantum wells in microcavities.~\cite{Littlewood2004mce}
They are interesting in their own right but counterflow superconductivity therein is impossible.

First, evidence for broken-symmetry phases conceptually similar to condensates of indirect excitons have been reported in GaAs CQW.~\cite{Eisenstein2004bec} These phases arise in the presence of a quantizing magnetic field at sub-Kelvin temperatures.
Next, a number of broken-symmetry states, some of which are similar to exciton condensates, have been predicted to form in bilayer graphene.~\cite{McCann2013epb}
The double-layer graphene (DLG) systems must be specially mentioned because a high-temperature coherent state in DLG was theoretically discussed recently.
The fundamental obstacle to exciton condensation in DLG is that monolayer graphene is a zero-gap semimetal with a linear quasiparticle dispersion.
Although in such a system phases of weakly bound excitons may exist, the corresponding critical temperature is extremely sensitive to the effective strength of the electron-hole attraction. Calculation of the latter requires accounting for screening of the long-range Coulomb interaction as well as short-range correlation effects, both of which are challenging problems. Theoretical estimates of the critical temperature in DLG range from hundreds of Kelvin \cite{Min2008rts, Perali2013hts, Nielsen2014ess} to a few milli-Kelvin.~\cite{Kharitonov2008ese} We wish to stress that there is no room for such an enormous uncertainty in our proposal based on MoS$_2$, a semiconductor with a significant bandgap, modest dielectic constant, and exceptionally stable exciton state.
The dependence of the characteristic temperatures $T_d$ and $T_{\mathrm{BKT}}$ on electron density is linear rather than exponential. All pertinent numerical factors are constrained by numerous prior studies of similar semiconducting systems.
Therefore, the error in the estimated temperatures should be small for the system considered in our paper.

The scaling law of Eq.~\eqref{eqn:T_d_max} elucidates a general principle for realizing high-temperature coherent states of excitons. The proposed design is an initial blueprint and is amenable to further optimization. For example, one~\cite{Komsa2013eso} or both of MoS$_2$ layers can be substituted by a different transition metal dichalcogenide (TMD), such as WS$_2$ or WSe$_2$.
The two MoS$_2$ monolayers may be replaced by a single several-monolayer-thick TMD encapsulated by hBN in which indirect excitons would be composed of electrons and holes confined at the opposite sides of the TMD layer.
The outlined principle for realizing high-temperature superfluidity can also be extended to layered materials other than TMD and hBN.

Besides providing a new platform for exploring fundamental quantum phenomena, indirect excitons in van der Waals structures can be also utilized for the development of optoelectronic circuits.~\cite{High2008cef}
In such devices
in-plane potential landscapes for excitons are created and controlled by external electric fields that couple to the permanent dipole moment $p = e c$ of the indirect excitons.
The operation temperature for excitonic
circuits in the van der Waals structures is expected to exceed by an order of magnitude the $\sim 100\,\mathrm{K}$ record \cite{Grosso2009eso} set by GaAs-based CQWs.

Finally, we note that Fig.~\ref{fig:T} was constructed following the example of the Monte-Carlo simulations~\cite{DePalo2002eci, Schleede2012pdb, Maezono2013eab} in which only exciton phases with two possible spin flavors were considered.
In TMDs, spin and orbital degeneracies of indirect excitons may have a more intriguing structure.
Such degeneracies can be controlled by strong spin-orbit coupling,~\cite{Kormanyos2014soc} polarization of the excitation beam,~\cite{Xiao2011csv, Ross2013ecn} or many-body interactions.~\cite{Ben-Tabou_de-Leon2003mtb}

Experimental realization of superfluidity and counterflow superconductivity as well as exciton circuits and spintronic/valleytronic devices in atomically thin heterostructures may have far-reaching implications for science and technology.

\small
\subsection*{Methods}
\paragraph{Interlayer tunneling.}
The action $S$ for tunneling of quasiparticles across the hBN spacer can be estimated from the usual formula for the rectangular potential barrier of height $U_b$:
\begin{equation}
S = 2 (c_b / a_B) (m_b / m_0) U_b^{1 / 2}(\mathrm{Ry})\,.
\label{eqn:S}
\end{equation}
Here $c_b = N c_1$ is the tunneling length, $N$ is the number of hBN layers, $c_1 = 0.333\,\mathrm{nm}$ is the thickness of one hBN layer, $a_B = 0.0529\,\mathrm{nm}$ is the hydrogen Bohr radius, and $m_b$ is the effective carrier mass inside the barrier.
If the Fermi level is in the middle of the hBN energy gap, we expect $U_b \approx 3\,\mathrm{eV}$.
Assuming also $m_b \approx 0.5$ in units of the bare electron mass, we obtain $S \approx 5 N$, which is in agreement with $S = 4.6 N$ deduced from the tunneling conductance measured in graphene/hBN/graphene structures.~\cite{Britnell2012fet}
In comparison, in GaAs/AlGaAs CQW structures with $m_b \approx 0.35$, $U_b \approx 0.15\,\mathrm{eV}$, and $d_b = 4.0\,\mathrm{nm}$ we get $S = 9.4$. Hence,
for $N = 2$ the tunneling rate is comparable to that in the GaAs CQW.~\cite{High2008cef, High2012sci}
More detailed estimates would have to include the trapezoidal shape of the barrier (Fig.~\ref{fig:device}b) and precise nature of the band alignment in the TMD/hBN structures.
Nevertheless, it is clear that by a minor adjustment of $N$ in the range from, say, $2$ to $4$, 
sufficiently
long interlayer tunneling lifetimes can be achieved for indirect excitons.

\paragraph{Binding energy of indirect excitons.}
We model both the electron and the hole that compose an indirect exciton as 2D quantum particles confined in the mid-planes of two separate MoS$_2$ layers.
The Schr\"odinger equation for the relative motion reads
\begin{equation}
-\frac{\hbar^2}{2 \mu} \nabla^2 \phi(\mathbf{r}) + U(\mathbf{r}) \phi(\mathbf{r}) = -E_{\mathrm{ind}} \phi(\mathbf{r})\,,
\label{eqn:Schroedinger}
\end{equation}
where $\mu$ is the reduced mass.
We model the potential $U(\mathbf{r})$ of electron-hole Coulomb interaction using the continuum-medium electrostatics.~\cite{Berkelbach2013tnc}
This simplified approach neglects frequency dependence of the dielectric functions of the materials involved.
We approximate each of MoS$_2$ layers as a uniaxial dielectric slab of thickness $c_2 = 0.312\,\mathrm{nm}$ with principal dielectric tensor components~\cite{Berkelbach2013tnc} $\epsilon_2^{\perp} = 14.29$ and $\epsilon_2^{\parallel} = 6.87$ in the directions perpendicular and parallel to the $z$-axis, respectively.
In turn, the hBN spacer is modelled as a slab of thickness $N c_1$, $c_1 = 0.333\,\mathrm{nm}$, with the dielectric constants~\cite{Cai2007irs} $\epsilon_1^{\perp} = 6.71$ and $\epsilon_1^{\parallel} = 3.56$.
We define the anisotropy parameter $\kappa_j$ and the effective dielectric constant $\epsilon_j$ of each material by
\begin{equation}
\kappa_j = {\sqrt{\epsilon_j^{\perp}}}\,\Big/\,{\sqrt{\epsilon_j^{\parallel}}}\,,
\quad
\epsilon_j = \sqrt{\epsilon_j^{\perp}}\, \sqrt{\epsilon_j^{\parallel}}\,,
\quad j = 1, 2\,,
\label{eqn:kappa_and_epsilon}
\end{equation}
so that $\kappa_2 = 1.44$, $\epsilon_2 = 9.91$ (MoS$_2$) and $\kappa_1 = 1.37$, $\epsilon_1 = 4.89$ (hBN).
The interaction potential can be found by solving the Poisson equation for this layered system, which is done by reducing it to a set of linear equations for the Fourier component $\tilde{U}(\mathbf{q})$ and using the inverse Fourier transform:
\begin{equation}
U(\mathbf{r}) = \int \frac{d^2 q}{(2\pi)^2}
 e^{i \mathbf{q} \mathbf{r}} \tilde{U}(\mathbf{q})\,.
\label{eqn:U_from_tildeU}
\end{equation}
By symmetry, $\tilde{U}(\mathbf{q})$ depends only on the magnitude $q = |\mathbf{q}|$ of vector $\mathbf{q}$ not on its direction.
After simple algebra, we arrive at the result $\tilde{U}(q) = C(q) / D(q)$, where functions $C(q)$ and $D(q)$ are given by
\begin{align}
C(q) &= \frac{8\pi e^2}{q}\, \epsilon_1
             \left[(\epsilon_1 - \epsilon_2) e^{-c_2 k_2 / 2} - (\epsilon_1 - \epsilon_2)e^{c_2 k_2 / 2}\right]^2 e^{-c k_1} ,
\label{eqn:C}\\
D(q) &= \left[(\epsilon_1 - \epsilon_2)^2 e^{-c_2 k_2}
            - (\epsilon_1 + \epsilon_2)^2
               e^{c_2 k_2}\right]^2 \notag\\
     &- 4 \left(\epsilon_1^2 - \epsilon_2^2 \right)^2 e^{-2 (c - c_2) k_1}
       \sinh^2 c_2 k_2\,.
\label{eqn:D}
\end{align}
Here we introduced the short-hand notations $c = N c_1 + c_2$ (the center-to-center distance) and $k_j = \kappa_j q$ (the $z$-direction wavenumber of the evanescent Fourier harmonics $e^{i q r \pm k_j z}$ in medium $j$).
From numerical calculations using Eqs.~\eqref{eqn:U_from_tildeU}--\eqref{eqn:D} we found that potential $U(r)$ is accurately approximated by the following analytical expression:
\begin{equation}
U(r) = \frac{e^2}{\epsilon_1}\, \frac{1}{\sqrt{r^2 + d^2}}
\left(1 - \frac{A}{1 + B r^2}\right)\,,
\quad d = \kappa_1 c\,.
\label{eqn:U}
\end{equation}
With a suitable choice of coefficients $0 < A < 1$ and $B > 0$ this form produces asymptotically exact results for $U(r)$ at both small and large $r$.

To solve Eq.~\eqref{eqn:Schroedinger}, we discretized it on a real-space 2D grid (typically, $75 \times 75$). The resultant linear eigenvalue problem was diagonalized by standard numerical methods yielding the binding energy $E_{\mathrm{ind}}(N)$ and the gyration radius $r_x(N)$ of indirect excitons. The latter is defined in terms of a normalized ground-state wavefunction $\phi(r)$ by means of the integral
\begin{equation}
r_x^2 = \int r^2 \phi^2(r) d^2 r\,.
\label{eqn:r_x}
\end{equation}
The results are shown in Fig.~\ref{fig:E_ind}. For $N = 2$ we find $r_x = 2.50\,\mathrm{nm} = 2.43\, a_x$ and $E_{\mathrm{ind}} = 87\,\mathrm{meV} \approx 0.6\,\mathrm{Ry}_x$.
This binding energy is an order of magnitude larger than $E_{\mathrm{ind}} = 4$--$10\,\mathrm{meV}$ typical for excitons in GaAs/AlGaAs CQW structures.~\cite{Szymanska2003ebc, Grosso2009eso}

\begin{figure}[h]
\begin{center}
\includegraphics[width=2.20in]{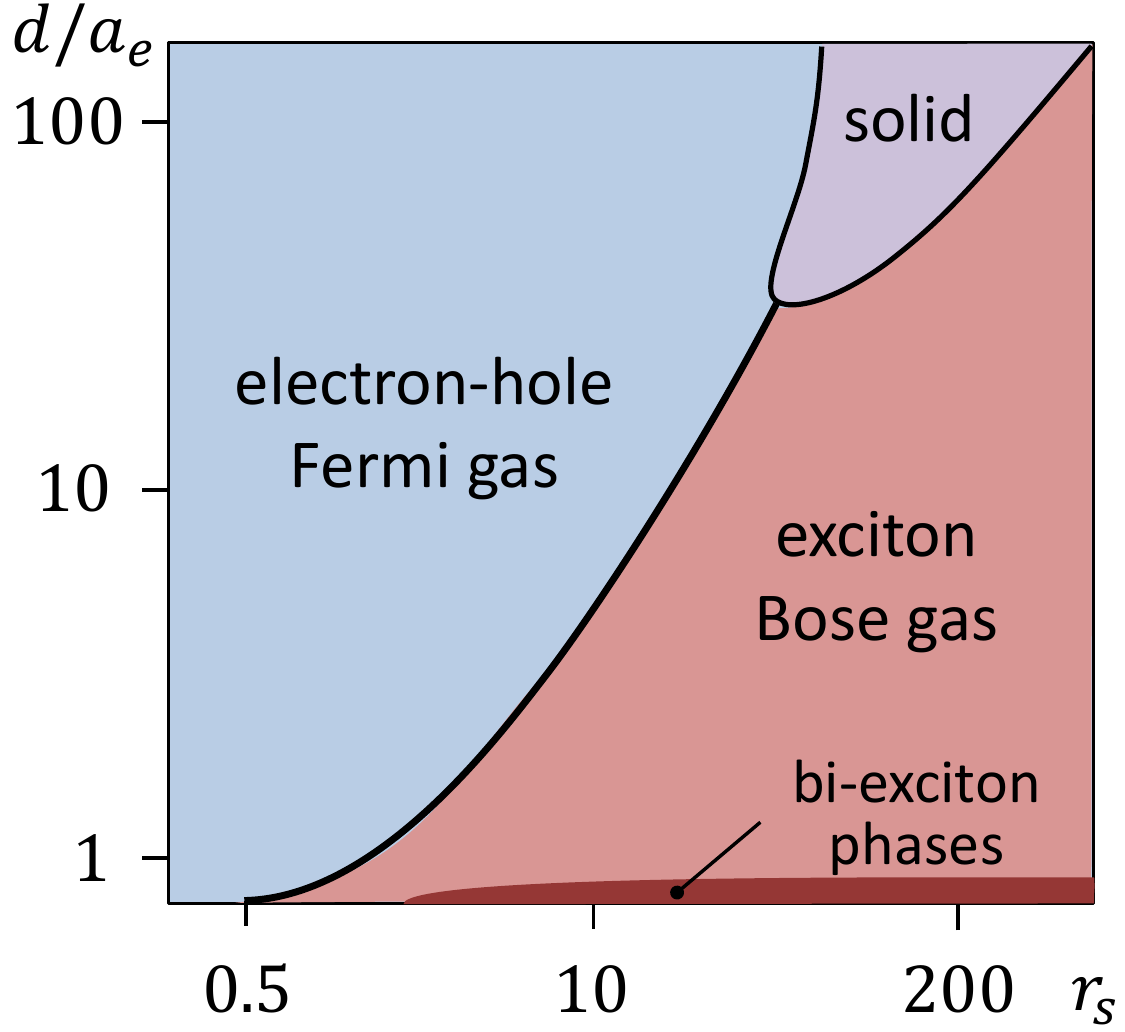}
\end{center}
\caption{\textbf{A schematic $\bm{T = 0}$ phase diagram.}
The system is an electron-hole bilayer in which the particles have spin $1/2$ but no valley degeneracy.
The MoS$_2$/hBN device with the $N = 2$ layer hBN spacer corresponds to $d / a_e \approx 2.8$.
At such $d / a_e$ the ground state of the system at $r_s \gtrsim 6$ is the superfluid Bose gas of indirect excitons.
\label{fig:0}
}
\end{figure}

\paragraph{Zero-temperature phases.}
It is instructive to complement the above discussion of the finite-$T$ phase diagram (Fig.~\ref{fig:T}) with commenting on
the $T = 0$ phases.
Such phases include electron-hole Fermi gas, exciton Bose gas, and exciton solid. The approximate phase boundaries
based on available Monte-Carlo calculations~\cite{DePalo2002eci, Astrakharchik2007qpt, Buechler2007sc2, Schleede2012pdb, Maezono2013eab} are shown in Fig.~\ref{fig:0}.
Although those simulations were done for electron-hole bilayers in vacuum, an approximate correspondence with our dielectric environment can be achieved if we neglect the correction term in the parenthesis in Eq.~\eqref{eqn:U}.
The mapping is then obtained by setting the distance between the layers to $d = \kappa_1 c$ and including the effective dielectric function $\epsilon_1$ in the definition of the electron Bohr radius:
\begin{equation}
a_e = \frac{\hbar^2 \epsilon_1}{m_e e^2} = \frac12 \frac{\hbar^2 \epsilon_1}{\mu e^2} = \frac{a_x}{2}\,.
\label{eqn:a_e}
\end{equation}
On the horizontal axis in Fig.~\ref{fig:0} we plot the dimensionless intra-layer distance parameter $r_s = 1 / \sqrt{\pi n_e a_e^2}$,
where $n_e$ is the total electron density.
Solid phases form at large $r_s$ and $d / a_e$.
The exciton dipole solid crosses over to the interlocked Wigner crystals of electrons and holes as the interlayer distance becomes larger than the characteristic intralayer one, $d \gtrsim r_s a_e$.

The considered MoS$_2$/hBN heterostructure with $N = 2$ corresponds to $d / a_e \approx 2.8$ for which the ground-state is never a solid phase. Instead, at large $r_s$, i.e., at low electron density, the ground state of the system is an exciton gas.
As $r_s$ decreases, the Mott transition to electron-hole Fermi gas occurs. At this transition the excitons dissociate due to screening and phase space filling.~\cite{Keldysh1968cpe}
The latest estimates~\cite{Maezono2013eab} of the $T = 0$ Mott transition at $d = 2.8 a_e$ give $r_s \approx 6$, which corresponds to Eq.~\eqref{eqn:n_M}.
Interestingly, a Bardeen-Cooper-Schrieffer-like excitonic state of a dense electron-hole gas predicted by the earlier theory~\cite{Keldysh1965pis} was not found in the cited Monte-Carlo calculations.~\cite{DePalo2002eci, Maezono2013eab}

Unless the effective inter-layer separation $d$ is small, the interaction between indirect excitons is dominated by the classical dipole repulsion term.~\cite{Zhu1995ecs, Lozovik1996pti}
However, as shown previously,~\cite{Meyertholen2008bit, Tan2005ebe, Schindler2008aee} at $d < 0.87\, a_e$ quantum exchange-correlation effects cause the change of repulsion to attraction leading to appearance of bi-excitons phases.~\cite{Lozovik1996pti, Maezono2013eab}
Narrow ranges of modulated phased (stripes, bubbles, or supersolids) may exist near any of the first-order phase transition lines.~\cite{Spivak2004pib}

%%%%% FIG %%%%%
%%%%%%%%%%%%%%%

\normalsize
%%%%%%%%%%%%%%%%%%%%%%%%%%%%%%%%%%%%%%%%%%%%%%%%%%%%%%%%%%%%%%%%%%%%%%%%%
%\bibliography{2DACX}

%%%%%%%%%%%%%%%%%%%%%%%%%%%%%%%%%%%%%%%%%%%%%%%%%%%%%%%%%%%%%%%%%%%%%%%%%% Acknowledgements
%%%%%%%%%%%%%%%%%%%%%%%%%%%%%%%%%%%%%%%%%%%%%%%%%%%%%%%%%%%%%%%%%%%%%%%%%
\subsection*{Acknowledgements}
{\small
This work is supported by US ONR \& UCOP (M.M.F), NSF (L.V.B), and
also European Research Council and EC-FET European Graphene Flagship (K.S.N.).
We are grateful to Andr\'e~K.~Geim for comments.
}
\subsection*{Author contributions}
{\small
All the authors contributed to the ideas, discussion, and writing of the paper. M.M.F. carried out numerical simulations
}
\subsection*{Additional information}
{\small
\paragraph{Competing financial interests:}
The authors declare no competing financial interests.

\paragraph{Reprints and permission} information is available online at \url{http://npg.nature.com/reprintsandpermissions/}

\paragraph{How to cite this article:} Fogler, M.~M., Butov, L.~V., and Novoselov, K.~S. High-temperature superfluidity with indirect excitons in van der Waals heterostructures. \textit{Nat. Commun.} 5:4555 (2014) \url{http://dx.doi.org/10.1038/ncomms5555}.
}
\end{document}